\documentstyle[epsf,11pt]{article}

\title{Dimensions of fractals related to languages defined by 
tagged strings in complete  genomes\thanks{This work was partially supported
by Chinese Natural Science Foundation and Chinese Postdoctoral Science Foundation.}}
\author{Zu-Guo Yu$^{1,2}$, Bai-lin Hao$^{2}$, Hui-min Xie$^{3,2}$ and Guo-Yi Chen$^{2}$\\
   {\small $^1$Department of Mathematics, Xiangtan University, Hunan 411105, P.R. China.}\\
   {\small $^{2}$Institute of Theoretical Physics, Academia Sinica},\\
     {\small P.O. Box 2735, Beijing 100080, P.R. China. }\\
     {\small $^3$Department of Mathematics, Suzhou University, Jiangsu 215006, P.R. China.}
  }
\date{}
\begin{document}
\maketitle
\newcommand{\be}{\begin{equation}}
\newcommand{\ee}{\end{equation}}
\newcommand{\ben}{\begin{eqnarray}}
\newcommand{\een}{\end{eqnarray}}
\newtheorem{Theorem}{\quad Theorem}[section]
\newtheorem{Proposition}{\quad Proposition}[section]
\newtheorem{Definition}{\quad Definition}[section]
\newtheorem{Lemma}{\quad Lemma}[section]
\newtheorem{Corollary}{\quad Corollary}[section]
\newtheorem{Example}{\quad Example}[section]

\begin{abstract}
  A representation of frequency of strings of length $K$ in complete  genomes
   of 
 many organisms in a square has led to seemingly self-similar patterns when $K$
 increases. These patterns are caused by under-represented strings with a certain
 ``tag"-string and they define some fractals in the $K\rightarrow\infty$ limit.
 The Box and Hausdorff dimensions of the limit set 
  are discussed. Although the method proposed by Mauldin and  Williams    
to calculate Box and Hausdorff dimension
is valid in our case, a different and
sampler method is proposed in this paper.
\end{abstract}


\section{Introduction}
 
\ \ \  In the past decade or so there has been a ground swell of interest in
 unraveling the mysteries of DNA. 
The heredity information of organisms (except for so-called RNA-viruses)
is encoded in their DNA sequence
which is a one-dimensional unbranched polymer made of four different
kinds of monomers (nucleotides): adenine ($a$), cytosine ($c$), guanine ($g$),
and thymine ($t$). As long as the encoded information is concerned we can
ignore the fact that DNA exists as a double helix of two ``conjugated''
strands and only treat it as a one-dimensional symbolic sequence made
of the four letters from the {\it alphabet} $\Sigma=\{a, c, g, t\}$. Since the
first complete genome of a free-living
bacterium {\it Mycoplasma genitalium} was sequenced in 1995$^{\cite{Fraser}}$,
 an ever-growing
number of complete genomes has been deposited in public databases.
The availability of complete genomes opens the possibility to
ask some global questions on these sequences. One of the simplest conceivable
questions consists in checking whether there are short strings of letters that
are absent or under-represented in a complete genome. The answer is in the
affirmative and the fact may have some biological meaning$^{\cite{hlz98}}$.

The reason why we are interested in absent or under-represented strings
is twofold. First of all, this is a question that can be asked only nowadays
when complete genomes are at our disposal. Second, the question makes sense as
one can derive a {\it factorizable} language from a complete genome which
would be entirely defined by the set of forbidden words.

   We start by considering how to visualize the avoided and under-represented
strings in a bacterial genome whose length is usually the order of a million
letters.

   Bai-lin Hao $^{\cite{hlz98}}$ {\it et al}. proposed a simple visualization
method based on counting and coase-graining the frequency of appearance of 
strings of a given length. When applying the method to all known complete genomes,
fractal-like patterns emerge. The fractal dimensions are basic and important 
quantities to characterize the fractal. One will  naturally ask the question:
what are  the fractal dimensions of the fractals rerlated to languages defined 
by tagged strings?
In this paper we will answer the question.    

\begin{figure}[h]
\centerline{\epsfxsize=5cm \epsfbox{ 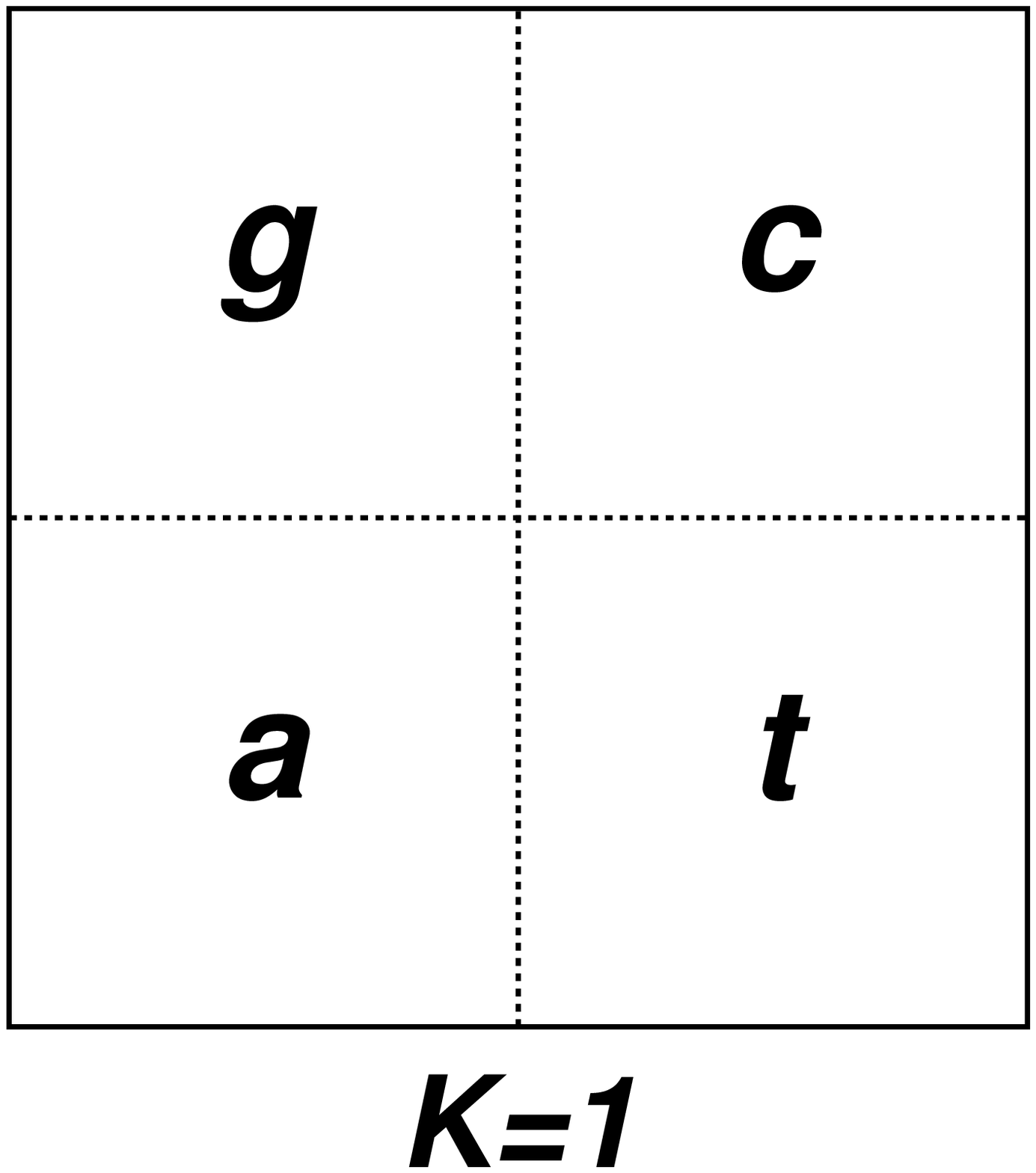}
\epsfxsize=5cm \epsfbox{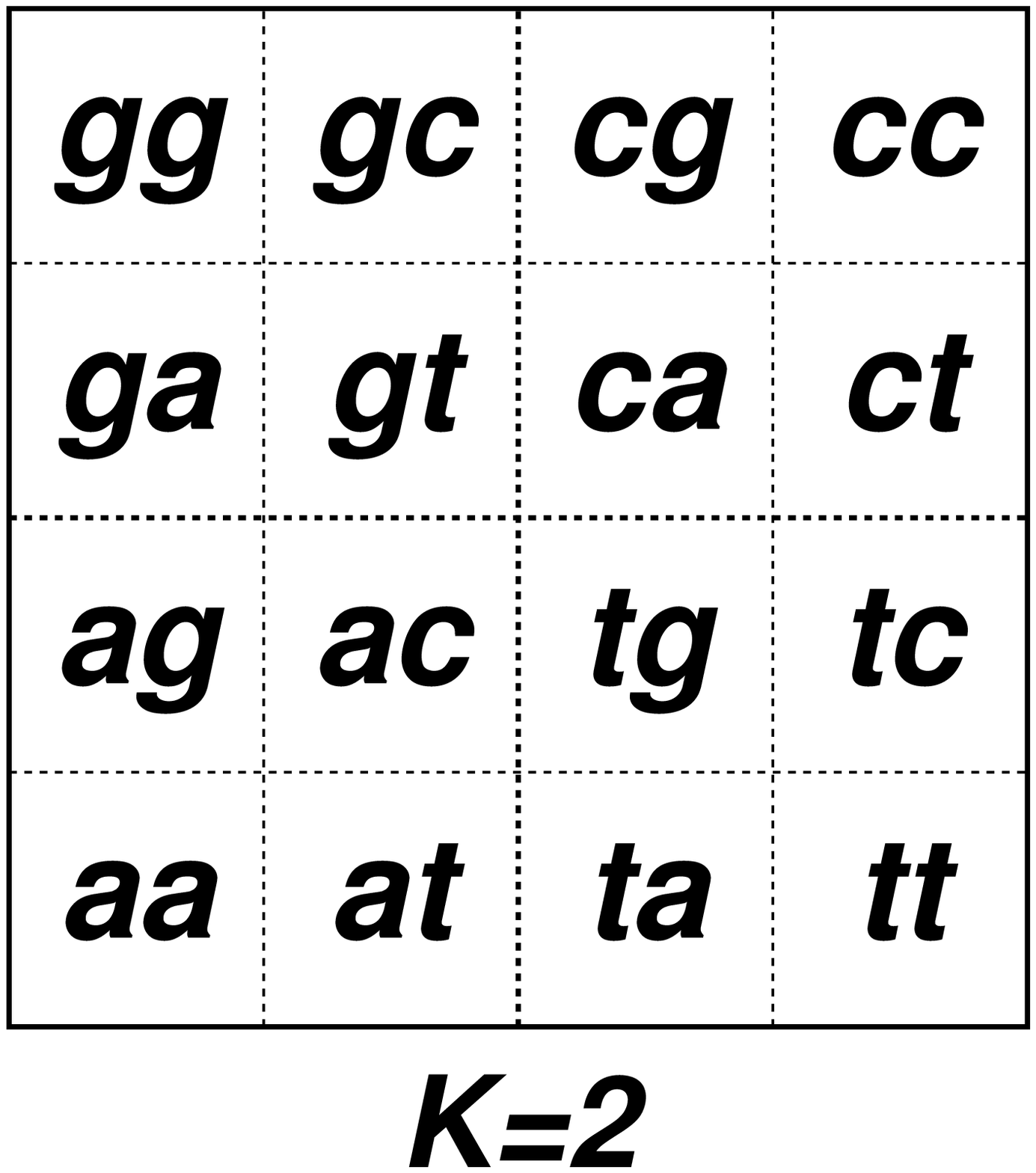}
\epsfxsize=5cm \epsfbox{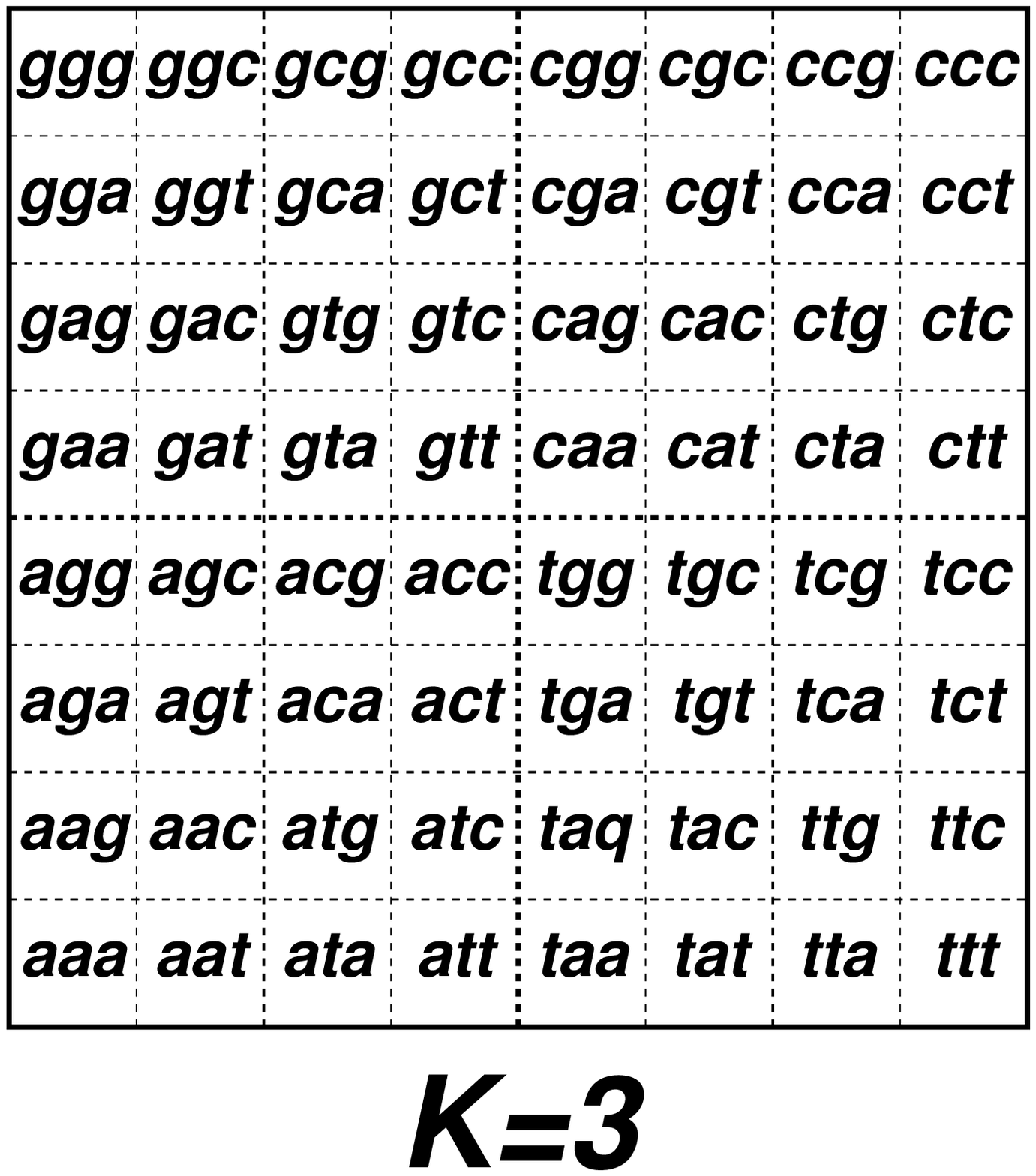}}
\caption{The arrangement of string counters for $K=1$ to~3 in squares
of the same size.}
\label{f1}
\end{figure}

\section{Graphical representation of counters}
\label{s3}

\ \ \ We call any string made of $K$ letters from the set $\{g, c, a, t\}$ a
$K$-string. For a given $K$ there are in total $4^K$ different $K$-strings.
In order to count the number of each kind of $K$-strings in a given DNA
sequence $4^K$ counters are needed. These counters may be arranged as a
$2^K\times 2^K$ square, as shown in Fig.~\ref{f1} for $K=1$ to~3.

\bigskip

\begin{figure}[h]
\centerline{\epsfxsize=6cm \epsfbox{ 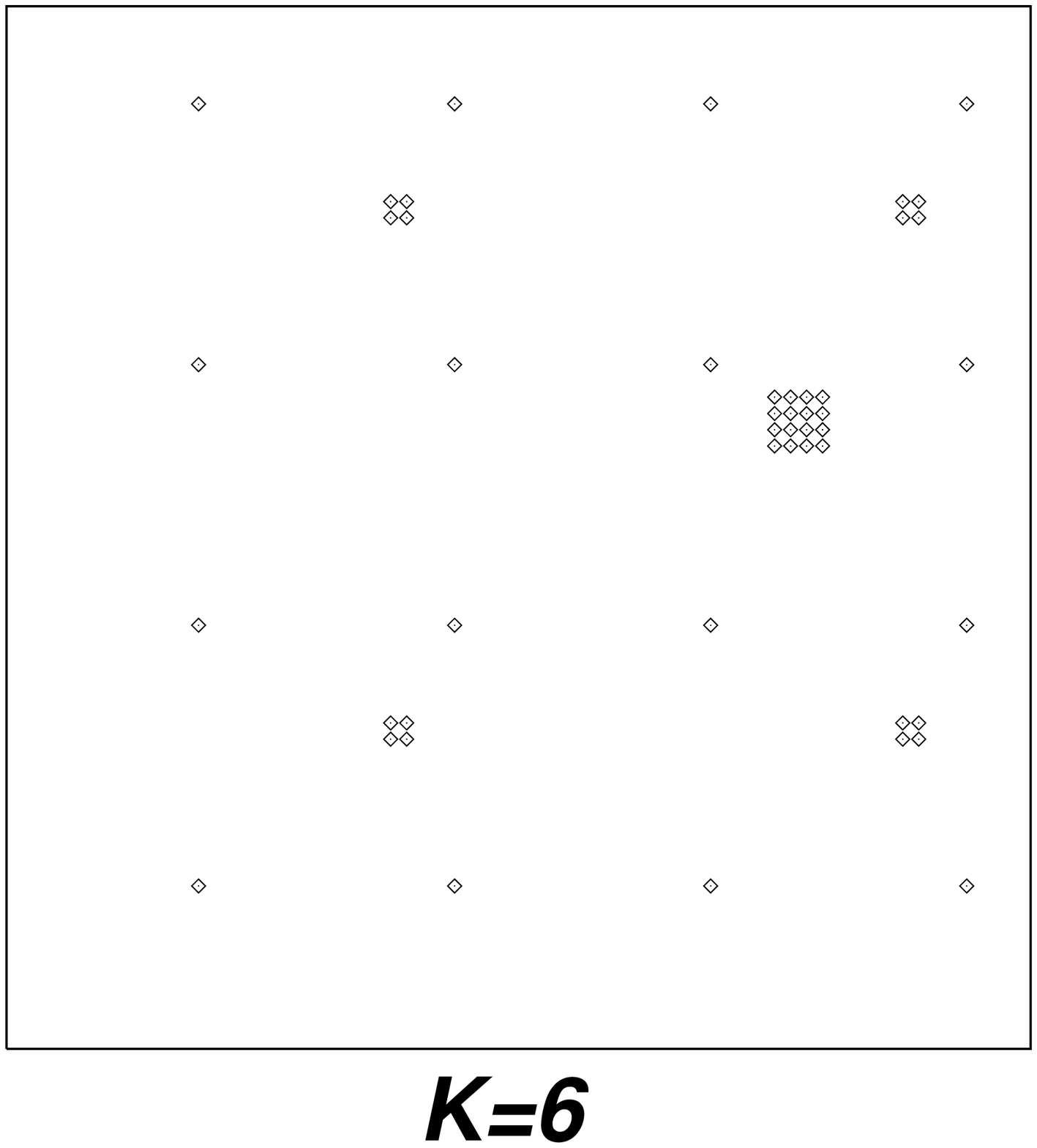}
\epsfxsize=6cm \epsfbox{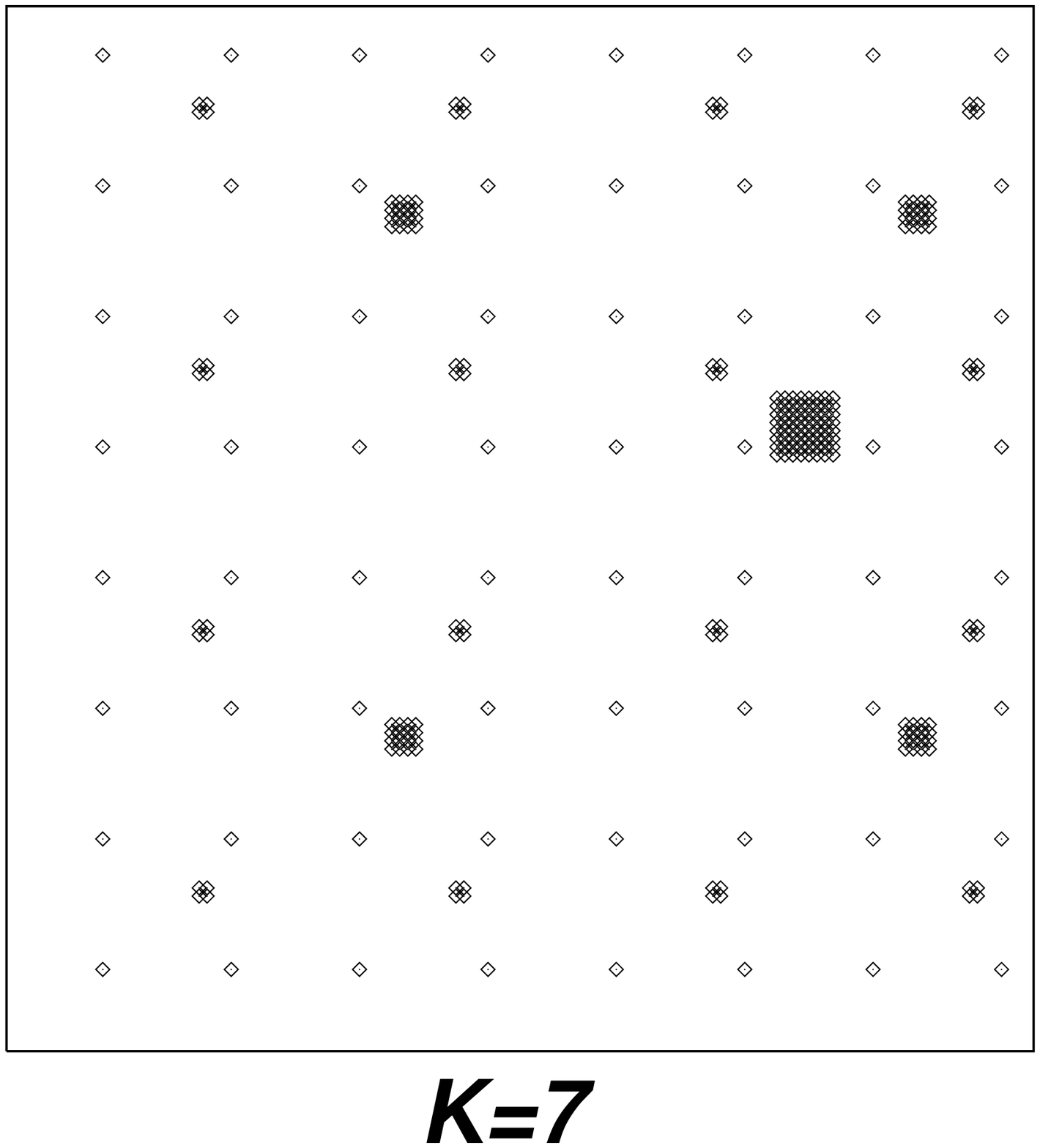}}
\centerline{\epsfxsize=6cm \epsfbox{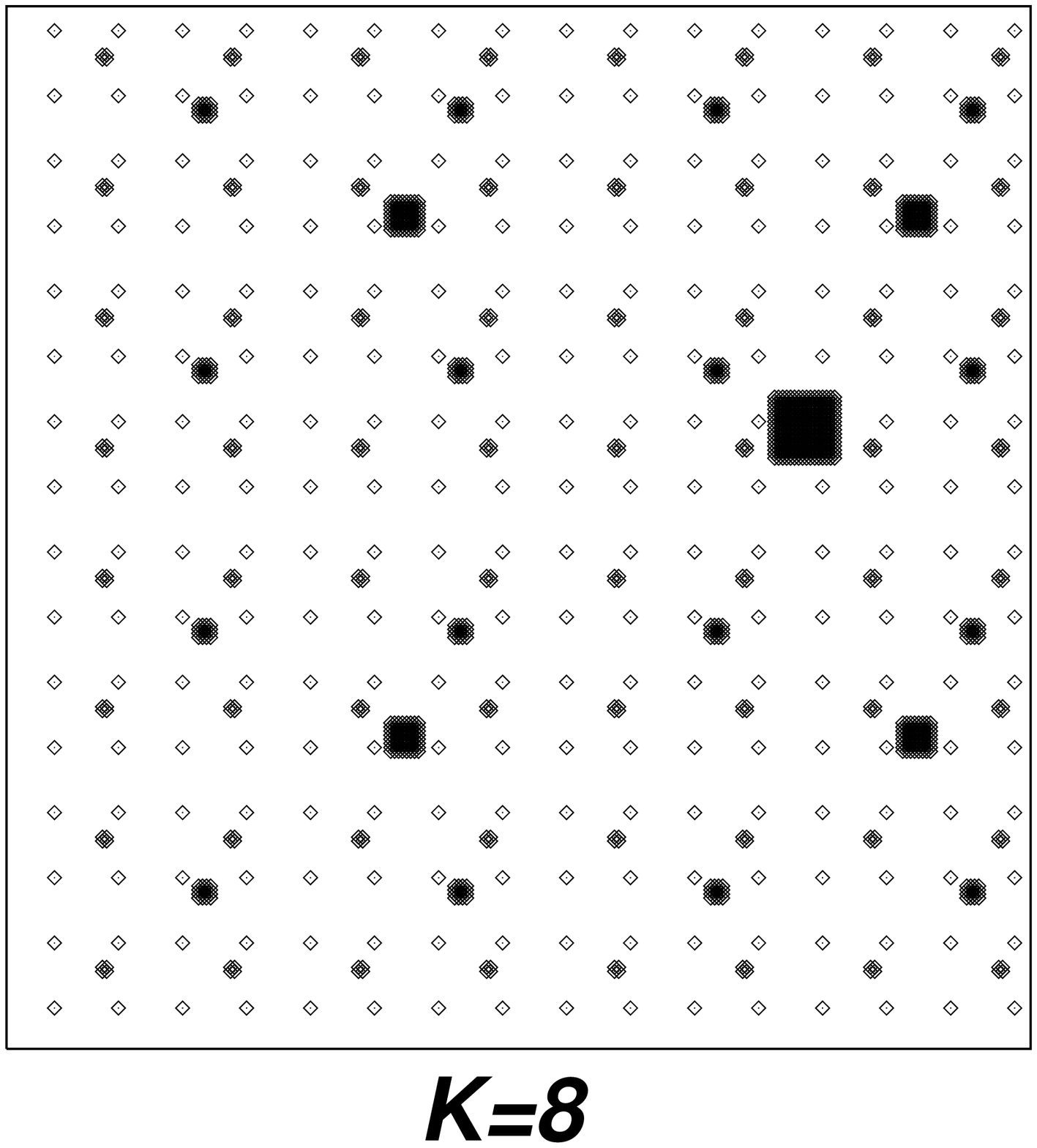}
\epsfxsize=6cm \epsfbox{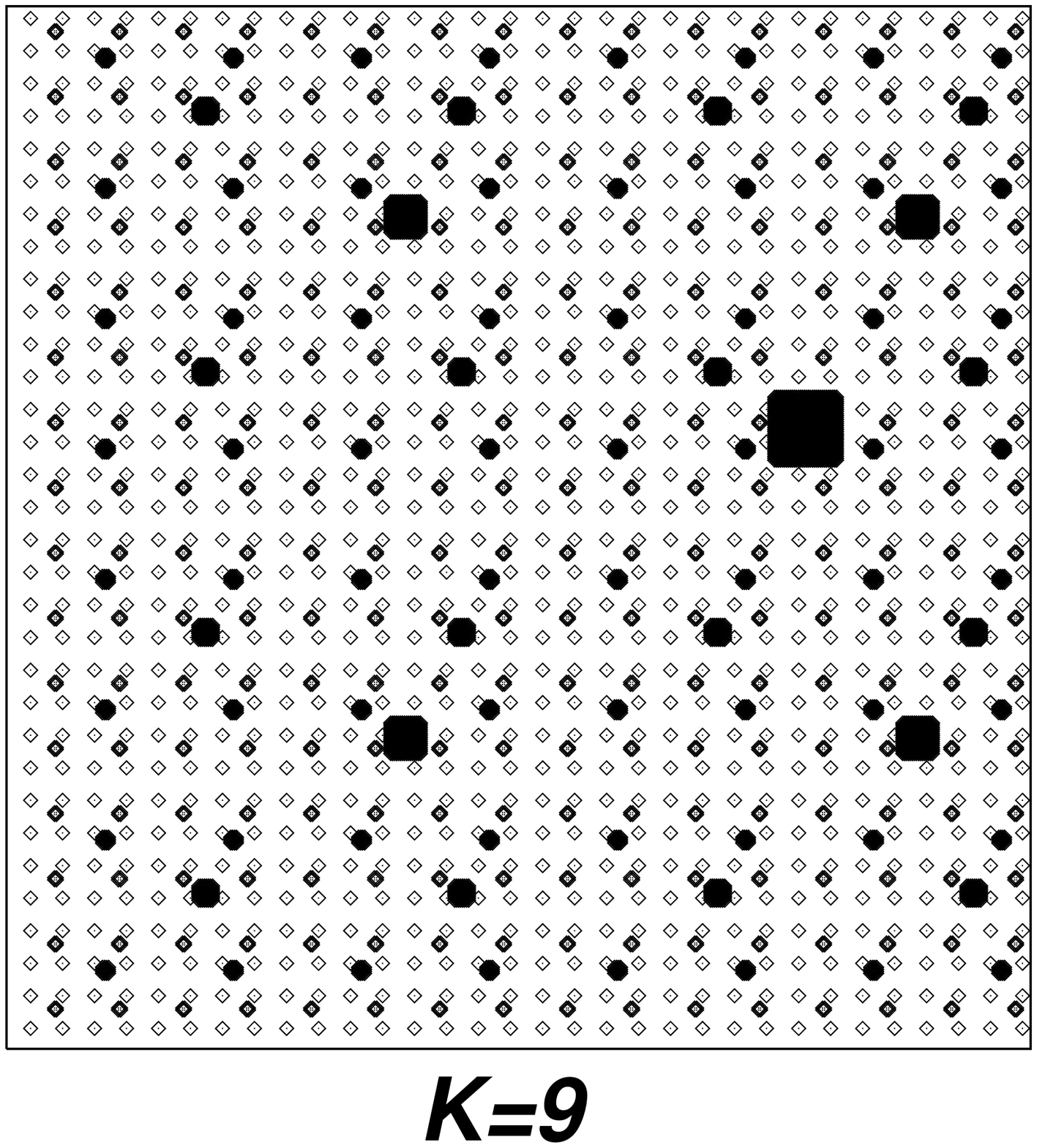}}
\caption{$ctag$-tagged strings in $K=6$ to 9 frames.}
\label{fig1}
\end{figure}

\bigskip
In fact, for a given $K$ the corresponding square may be represented as a
direct product of $K$ copies of identical matrices:
$$ M^{(K)}=M\otimes M\otimes \cdots \otimes M,$$
where each $M$ is a $2\times 2$ matrix:
$$M=\left[\begin{array}{cc} g & c \\ a & t\\ \end{array}\right],$$
which represents the $K=1$ square in Fig.~\ref{f1}. For convenience of
programming, we use binary digits 0 and 1 as subscripts for the matrix
elements, i.e., let $M_{00}=g$, $M_{01}=c$, $M_{10}=a$, and $M_{11}=t$.
The subscripts of a general element of the $2^K\times 2^K$ direct product
matrix $M^{(K)}$,
$$ M^{(K)}_{I,J}=M_{i_1j_1}M_{i_2j_2}\cdots M_{i_Kj_K}$$
are given by $I=i_1i_2\cdots i_K$ and $J=j_1j_2\cdots j_K$. These
may be easily calculated from an input DNA sequence
$$ s_1s_2s_3\cdots s_Ks_{K+1}\cdots,$$
where $s_i\in \{g, c, a, t\}$.
We call this $2^K\times 2^K$ square a $K$-frame.
 Put in a frame
of fixed $K$ and described by a color code biased towards small counts, each
bacterial genome shows a distinctive pattern which indicates on absent or
under-represented strings of certain types${}^{\cite{hlz98}}$. For example,
many bacteria avoid strings containing the string $ctag$. Any string that
contains $ctag$ as a substring will be called a $ctag$-tagged string. If we
mark all $ctag$-tagged strings in frames of different $K$, we get pictures as
shown in Fig.~\ref{fig1}. The large scale structure of these pictures persists
but more details appear with growing~$K$. Excluding the area occupied by these
tagged strings, one gets a fractal $F$ in the $K\rightarrow\infty$ limit. It is
natural to ask what are the fractal dimensions of $F$ for a given tag.

In fact, this is the dimension of the complementary set of the tagged strings.
The simplest case is that of $g$-tagged strings. As the pattern has an
apparently self-similar structure the dimension is easily calculated to be
$$ \dim_H(F)=\dim_B(F)=\displaystyle\frac{\log 3}{\log 2},$$
where $\dim_H(F)$ and $\dim_B(F)$ are the Hausdorff and Box
dimensions$^{\cite{Fal}}$ of $F$.

In formal language theory, we  starts with alphabet
$\Sigma=\{a, c, g, t\}$. Let $\Sigma^\ast$ denotes the collection of all
possible strings made of letters from $\Sigma$, including the empty string
$\epsilon$. We call any subset $L\subset \Sigma^\ast$  a {\it  language}
over the alphabet $\Sigma$.
Any string over $\Sigma$ is called a {\it word}.
If we denote the given tag as $w_0$, for our case, 
$$L=\{\hbox{word which does not contain}\ w_0 \ \hbox{as factor}\}.$$
 $F$ is called the fractal related to language $L$.

\section{Box dimension of fractals}

 \ \ \ When we discuss the Box dimension, we can consider more general case, i.e. the case of
 more than one tag. We denote the set of tags as $B$, and assume that there has not one element
 being factor of any other element in $B$. We define
  $$L_1=\{\hbox{word which does not contain any of element of }\ B\ \hbox{as factor}\} $$
 
   Now let $a_K$ be the number of all strings of length~$K$ that belong to language $L_1$.
    As the linear size $\delta_K$ in the $K$-frame is
$1/{2^K}$, the Box dimension of $F$ may be calculated as:
\be
\dim_B(F)=\lim_{K\rightarrow\infty}\frac{\log a_K}{-\log \delta_K}=
\lim_{K\rightarrow\infty}\frac{\log {a_K}^{1/K}}{\log 2}.\label{hxy1}
\ee
  Now we define the generating function of $a_K$ as
$$f(s)=\displaystyle\sum_{K=0}^\infty a_K s^K,$$
where $s$ is a complex variable.

   First $L_1$ is a dynamic language, form Theorem 2.5.2 of ref.\cite{xie96}, we have
    \be
     \lim_{K\rightarrow\infty}a_K^{1/K} \qquad \hbox{exists, we denote it as }\ l.  
\ee
From (\ref{hxy1}), we have
\be \dim_B(F)=\frac{\log l}{\log 2}.
\label{hxy2} \ee
   
   For any word $w=w_1w_2\dots w_n,w_i\in \Sigma$ for $i=1,\dots,n$, we denote
\begin{eqnarray*}
 Head(w)&=&\{w_1,\ w_1w_2,\ w_1w_2w_3,\ \dots,\ w_1w_2\dots w_{n-1}\},\\
 Tail(w)&=&\{w_n,\ w_{n-1}w_n,\ w_{n-2}w_{n-1}w_n,\ \dots,\ w_2w_3\dots w_n\}.
 \end{eqnarray*}
 For given two words $u$ and $v$, we denote $overlap(u,v)=Tail(u)\cap Head(v)$.
 If $x\in Head(v)$, then we can write $v=xx'$. We denote $x'=v/x$ and define
 $$u:v=\sum_{x\in overlap(u,v)} s^{|v/x|},$$
 where $|v/x|$ is the length of word $v/x$.
From Golden-Jackson Cluster method$^{\cite{nz98}}$, we can know that
$$ f(s)=\frac{1}{1-4s-weight({\cal C})},$$
where $weight({\cal C})=\sum_{v\in B}weight({\cal C}[v])$ and 
$weight({\cal C}[v])$ ($v\in B$)  are solutions of the linear equations:
$$weight({\cal C}[v])=-s^{|v|}-(v:v) weight({\cal C}[v])-\sum_{
\stackrel{u\in B}{u\neq v}}(u:v) weight({\cal C}[u]).$$
It is easy to see that $f(s)$ is a rational function. Its maximal analytic disc at
center 0 has radius $|s_0|$, where $s_0$ is the minimal module zero point of $f^{-1}(s)$.
On the other hand,  
 according to the Cauchy criterion of convergence we have
$1/l$ is the radius of convergence of series expansion of $f(s)$. Hence 
 $|s_0|=1/l$. From (\ref{hxy2}), we obtain the following result.
 
 \begin{Theorem} The Box dimension of $F$ is 
$$ \dim_B(F)=-\frac{\log|s_0|}{\log 2},
$$
where $s_0$ is the minimal module zero point of $1/f(s)$ and $f(s)$ is the 
generating function of language $L_1$.
\label{th3.3.1}
\end{Theorem}

In particular, the case of a single tag ---$B$ contains only one word --- is 
easily treated and some of the results are shown in Table~\ref{t1}.

\begin{table}[htb]
\begin{center}
\begin{tabular}{ccc|ccc}
\hline
Tag & $f(s)$ & $D$ & Tag & $f(s)$ & $D$\\
\hline
$g$   & $\frac{1}{1-3s}$ & $\frac{\log 3}{\log 2}$ &
$ggg$ & $\frac{1+s+s^2}{1-3s-3s^2-3s^3}$& 1.98235\\[0.2cm]
$gc$  & $\frac{1}{1-4s+s^2}$& 1.89997 &
$ctag$& $\frac{1}{1-4s+s^4}$& 1.99429\\[0.2cm]
$gg$  & $\frac{1+s}{1-3s-3s^2}$& 1.92269 &
$ggcg$& $\frac{1+s^3}{1-4s+s^3-3s^4}$ & 1.99438\\[0.2cm]
$gct$ & $\frac{1}{1-4s+s^3}$ & 1.97652 &
$gcgc$& $\frac{1+s^2}{1-4s+s^2-4s^3+s^4}$ & 1.99463\\[0.2cm]
$gcg$ & $\frac{1+s^2}{1-4s+s^2-3s^3}$& 1.978 &
$gggg$& $\frac{1+s+s^2+s^3}{1-3s-3s^2-3s^3-3s^4}$ & 1.99572\\
\hline
\end{tabular}
\end{center}
\caption{Generating function and dimension for some single tags.}
\label{t1}
\end{table}

\section{Hausdorff dimension of fractals}

 \ \ \  We obtained the Box dimension of $F$ in the previous section. Now 
  one will naturally ask whether the Hausdorff dimension of $F$ equals to the 
  Box dimension of it. In this section we will discuss the Hausdorff
  dimension of $F$. Now we only discuss the case of $B$ contains only one word $w_0$.
  From the $K$-frames ($K=|w_0|,|w_0|+1,\dots$), we can find:
  
  \begin{Proposition}
  $$\frac{\log 3}{\log 2}\leq \dim_H(F)\leq \dim_B(F)\leq \frac{\log (4^{|w_0|}-1)}{\log 2}<2.$$
 \end{Proposition}
 
 Now we denote $\alpha=-\frac{\log |s_0|}{\log 2}$ and $\alpha_K=\frac{\log a_K^{1/K}}{\log 2}$.
 
   For any word $w=w_1w_2\dots w_K$, we denote $F_{w_1w_2\dots w_K}$ the corresponding close
 square in $K$-frame and denote 
 $$F_K=\cap_{w=w_1w_2\dots w_K\in L}F_{w_1w_2\dots w_K},$$
  then $F=\lim_{K\rightarrow\infty}F_K$. 
  
  We first prove $\dim_H(F)=\dim_B(F)$ under a condition using elementary method.
  
  \begin{Lemma}: Suppose $E\subset {\bf R}^2$ with $|E|<1/2$, let
  \begin{eqnarray*}
   B_1=\{w=w_1w_2\dots w_K\in L:& \ & |F_{w_1w_2\dots w_K}|<|E|
   \leq |F_{w_1w_2\dots w_{K-1}}|\\
       & &\hbox{and}\ E\cap F_{w_1w_2\dots w_K}\neq\emptyset\},
   \end{eqnarray*}
   then $\# B_1\leq 2\pi$.
   \end{Lemma}
   {\it Proof}. Note that for each $w=w_1w_2\dots w_K\in B_1$
   $$\frac{|E|}{|F_{w_1w_2\dots w_K}|}\leq \frac{|F_{w_1w_2\dots w_{K-1}}|}{|F_{w_1w_2\dots w_K}|}=
  \frac{1}{2},$$
   then $|E|\leq \frac{1}{2}|F_{w_1w_2\dots w_K}|$.
   The interiors of $F_{w_1w_2\dots w_K}$ with $w=w_1w_2\dots w_K \in B_1$ are non-overlapping and all lie
   in a disc with radius $2|E|$, and all $F_{w_1w_2\dots w_K}$ are squares, hence
   $$(2|E|)^2\pi\geq (\frac{1}{\sqrt{2}}|F_{w_1w_2\dots w_K}|)^2\# B_1 \geq \frac{1}{2}(2|E|)^2\# B_1,$$
   hence $\# B_1\leq 2\pi$.

   $\Box$

   For any $w=w_1\dots w_{|w|},r\in\Sigma$, we denote $w\ast r=w_1\dots w_{|w|}r$
   and
   define $\nu_w=\nu_{w_1}\nu_{w_2}\dots\nu_{w_{|w|}}$, where
   $$\nu_{w_j}=\left\{\begin{array}{ll} 2^{\alpha}/4, & \quad \hbox{if}\ \#\{
   r\in \Sigma:\ w_1w_2\dots w_{j-1}r\in L\}=4,\\
   2^{\alpha}/3, & \quad \hbox{if}\ \#\{
   r\in \Sigma:\ w_1w_2\dots w_{j-1}r\in L\}=3. \end{array}\right. $$
   We assume $$
   (C_1)\qquad \nu_w=\nu_{w_1}\nu_{w_2}\dots\nu_{w_{|w|}}<M \ \hbox{(a constant)
   for any}\ w\in L.$$
   Now we have:

\begin{Theorem}
Under condition $(C_1)$, we have
$$\dim_H(F)=\dim_B(F)=\alpha \quad \hbox{and} \quad 0<{\cal H}^{\alpha}(F)<\infty,$$
 where ${\cal H}^{\alpha}(F)$ is the Hausdorff measure of $F$. \label{th3.4.1}
 \end{Theorem}

{\it Proof}.   We first prove
 that \be {\cal H}^{\alpha}(F)<\infty, \label{hxy3} \ee
 Since $\alpha_K\rightarrow \alpha$ as $K\rightarrow \infty$,
 for any small $\varepsilon>0$, there exists a integer $N>0$ such that for any $K>N$, we have $
 \alpha>\alpha_K-\varepsilon$. Hence
 \begin{eqnarray*}
 \sum_{w=w_1w_2\dots w_K \in L}|F_{w_1w_2\dots w_K}|^{\alpha}&=& a_K(\frac{1}{2})^{K \alpha}
 <a_K(\frac{1}{2})^{K(\alpha_K-\varepsilon)}\\
 &=&(\frac{1}{2})^{-K\varepsilon}\leq (\frac{1}{2})^{-(N+1)\varepsilon}<\infty.
 \end{eqnarray*}
 Hence ${\cal H}^{\alpha}(F)<\infty$.

    Now we want to prove ${\cal H}^{\alpha}(F)>0$.   We denote
   $$ \Sigma^{\infty}=\{\tau=\tau_1\tau_2\dots: \ |\tau|=\infty \ \hbox{and}\ \tau_1
   \dots \tau_K\in L \
   \hbox{for}\ K=1,2,\dots\}$$
   For any $\tau=\tau_1\tau_2\dots \ \in \Sigma^{\infty}$, we denote $\tau |_K
   =\tau_1\tau_2\dots \tau_K$,
   and define a probability measure $\widetilde{\mu}$ on $\Sigma^{\infty}$ by
   $$\widetilde{\mu}([w])=(\frac{1}{2})^{|w|\alpha}\nu_w,\quad \hbox{where}\
   [w]=\{\tau\in \Sigma^{\infty}:\ \tau|_{|w|}=w\}.$$
   We can see
   \begin{eqnarray*}
   \sum_{w\ast r\in L,r\in \Sigma}\widetilde{\mu}([w\ast r])&=&
   \sum_{w\ast r\in L,r\in \Sigma}(\frac{1}{2})^{(|w|+1)\alpha}\nu_{w\ast r}\\
   =(\frac{1}{2})^{|w|\alpha}\nu_w\sum_{w\ast r\in L,r\in \Sigma}(\frac{1}{2})
   ^{\alpha}\nu_r& =&(\frac{1}{2})^{|w|\alpha}\nu_w=\widetilde{\mu}([w]).
   \end{eqnarray*}
   There exists a natural continuous map $f$  from $\Sigma^{\infty}$ to $F$.
   Now we transfer $\widetilde{\mu}$ to a probability measure on $F$, let
   $\mu=\widetilde{\mu}\circ f^{-1}$. We will show that there is some constant
   $M_1>0$ such that if $E$ is a Borel subset of ${\bf R}^2$ with $|E|<1/2$, then
   $\mu(E)\leq M_1|E|^{\alpha}$. Of course, this inequality implies ${\cal H}^{\alpha}
   (F)\geq 1/M_1>0$.

   Set \begin{eqnarray*}
   B_1=\{w=w_1w_2\dots w_K\in L:& \ & |F_{w_1w_2\dots w_K}|<|E|\leq |F_{w_1w_2\dots w_{K-1}}|\\
       & &\hbox{and}\ E\cap F_{w_1w_2\dots w_K}\neq\emptyset\}.
   \end{eqnarray*}
Then
\begin{eqnarray*}
 \mu(E)&\leq&\sum_{w\in B_1}\widetilde{\mu}([w])\leq\#B_1
 |F_{w_1w_2\dots w_K}|^{\alpha}\nu_w\\
       &\leq&\#B_1|E|^{\alpha}\nu_w\leq 2\pi M |E|^{\alpha}=M_1|E|^{\alpha}.
\end{eqnarray*}

   $\Box$

\begin{Theorem}
If the length of tag $|w_0|\geq 3$ and for any $w \in L$, $\nu_w$ has the form
$$\nu_w=(\frac{2^{\alpha}}{3})(\frac{2^{\alpha}}{4})^{i_1}
(\frac{2^{\alpha}}{3})(\frac{2^{\alpha}}{4})^{i_2}(\frac{2^{\alpha}}{3})\cdots$$
or
$$\nu_w=(\frac{2^{\alpha}}{4})^{i_1}(\frac{2^{\alpha}}{3})
(\frac{2^{\alpha}}{4})^{i_2}(\frac{2^{\alpha}}{3})(\frac{2^{\alpha}}{4})^{i_3}\cdots$$
where  $i_1, i_2$ and $ i_3$ are positive integers, then $\dim_H(F)=\dim_B(F)=\alpha$
and $0<{\cal H}^{\alpha}(F)<\infty$.
\end{Theorem}
{\it Proof}. Since $|w_0|\geq 3$, we have $\alpha>\frac{\log 12}{2\log2}$, hence
$$(\frac{2^{\alpha}}{3})(\frac{2^{\alpha}}{4})>1.$$ Form the other condition,
we know that there exists $M_1=\max\{(\frac{2^{\alpha}}{3}),1\}$ such that
$\nu_w\leq M_1$ for any $w\in L$. Then from Theorem \ref{th3.4.1}, we can obtain
our result of this theorem.

$\Box$

{\bf Examples}: $w_0=ctg$ or $w_0=ctag$, the result $\dim_H(F)=\dim_B(F)$ holds.

If we do not have condition $(C_1)$, in the following we still can obtain  $\dim_H(F)=\dim_B(F)$.

We define $B_2=\{u\in \Sigma^{\ast}|\quad |u|=|w_0|,u\neq w_0\}$. One can know
the set $B_2$ contains $N_1=4^{|w_0|}-1$ elements, hence we can write 
$B_2=\{u_1,u_2,\dots ,u_{N_1}\}$. Now we can define a $N_1\times N_1$ matrix ${\cal A}$
by

$${\cal A}=[t_{i,j}]_{i,j\leq N_1},$$
where $t_{i,j}=(1/2)^\beta$ if $u_i=r_1x$ and $u_j=xr_2$ with $|x|=|w_0|-1, r_1,r_2\in \Sigma$,
and $t_{i,j}=0$ otherwise, and where $\beta$ satisfies $\Phi (\beta)=1$ with 
$\Phi(\beta)$ being the largest nonnegative eigenvalue of ${\cal A}$. Then from
the results of ref.\cite{MW88}, we have
\begin{Theorem}
If $B=\{w_0\}$, then 
$$\dim_H(F)=\dim_B(F)=\beta \quad \hbox{and}
 \quad 0<{\cal H}^{\alpha}(F)<\infty.$$
\end{Theorem}

From Theorem \ref{th3.3.1} and Theorem \ref{th3.4.1}, we have
\begin{Corollary}
If $B=\{w_0\}$, then $$\beta=\dim_H(F)=\dim_B(F)=\alpha .$$ \label{Co3.4.1}
\end{Corollary}

{\bf Remark}:
When $B$ contains more than one word, we can also construct a 
matrix ${\cal A}$ similarly, then from the results of ref.\cite{MW88}, we can
obtain the same conclusions of Theorem 4.3 and Corollary \ref{Co3.4.1} for this case.
From Corollary \ref{Co3.4.1},
we have two methods to calculate the Hausdorff and Box dimensions of $F$, i.e.
calculate $\alpha$ and $\beta$ respectively.

\section*{ACKNOWLEDGMENTS}
\ \ \   The first author would like to express his thanks to Prof. Zhi-Ying Wen for
 encouragement, and to Dr. Hui Rao and De-Jun Feng for 
many usful discussions.

\end{document}